\documentclass[12pt]{article}
\usepackage{setspace}
\usepackage{epsfig}
\usepackage{amssymb}
\def\be{\begin{equation}}
\def\ee{\end{equation}}
\def\ba{\begin{array}}
\def\ea{\end{array}}
\def\beqn{\begin{eqnarray}}
\def\eeqn{\end{eqnarray}}

\def\bt{\begin{tabular}}
\def\et{\end{tabular}}
\def\bc{\begin{center}}
\def\ec{\end{center}}
\begin{document}
\title{$~~~~$Investigating non-Fritzsch like texture specific quark mass matrices}
\author{Neelu Mahajan$^1$, Rohit Verma$^2$, Manmohan Gupta$^1$ \\
\\
{$^1$ \it Department of Physics, Centre of Advanced Study, P. U., Chandigarh, India.}\\
{$^2$ \it Rayat Institute of Engineering and Information Technology, Ropar, India.}\\
{\it Email: mmgupta@pu.ac.in}}

\maketitle
\begin{abstract}
A detailed investigation of all possible textures of Fritzsch-like
and non-Fritzsch like, 144 for texture 6 zero and 432 for texture 5 zero
 mass matrices, have been carried out to ascertain their compatibility
 with the existing quark mixing data. It seems that all the texture 6
zero possibilities are completely ruled out whereas in the case of
texture 5 zero mass matrices the only viable possibility looks to be
that of Fritzsch-like.
\end{abstract}

Texture specific mass matrices are known to provide satisfactory
 explanation of quark mixing phenomenon \cite{fritzsch}-\cite{t53}.
 In fact, it has been shown in literature that texture 6 zero
Fritzsch mass matrices are completely ruled out
 \cite{fritzsch,Ramond} while texture 5 zero Fritzsch-like mass matrices
may not be completely ruled out \cite{fritzsch}-\cite{t53}.
 Similarly, in the context of neutrino oscillation phenomenon,
intensive investigations have been carried out using texture
specific mass matrices \cite{xing}-\cite{neu8}. In case we have to
tackle the larger issue of quark and lepton mixing phenomena at a
deeper level, it is perhaps desirable to take into consideration
the quark lepton unification hypothesis \cite{qlU}. This
immediately brings forth the issue of finding the simplest texture
structures at the leading order, compatible with the quark and
lepton mixing phenomena. In view of absence of any theoretical
justification for Fritzsch-like mass matrices, it becomes
essential from the phenomenological point of view to consider
non-Fritzsch like mass matrices for quarks as well as neutrinos.
In the case of neutrinos, such an exercise has been done
\cite{xing},\cite{neu}-\cite{neu8}, however, a similar exercise
has not been carried
out for quarks, therefore it is desirable to carry out detailed
investigations for non-Fritzsch like quark mass matrices also.

Such an exercise is non trivial as can be seen by consideration of
texture 3 zero Fritzsch-like hermitian mass matrix given as
\be
 M=\left( \ba{ccc}
0 & A  & 0      \\ A^{*} & 0 &  B     \\
 0 &   B^{*}  & C \ea \right),
\ee\label{fm} where $A$ and $B$ are complex elements and $C$ is a
real element of the mass matrix. To obtain texture 6 zero mass
matrix we have to take mass matrix in the up sector as well as in
down sector of the above form.
 In case we do not restrict ourselves to this form one can
immediately find that there are large number of possibilities
 which are non-Fritzsch like as can be seen by distributing
non zero elements in the possible 9 slots available keeping in
mind the hermiticity of matrix. In the case of quark mixing
phenomenon, texture 6 zero Fritzsch quark mass matrices are
completely ruled out by quark mixing data \cite{fritzsch, Ramond},
whereas it has not been shown whether non-Fritzsch like mass
matrices are also ruled out. Similarly, in the case of texture 5
zero mass matrices, a detailed investigation of non Fritzsch-like
mass matrices has not been carried out.

The purpose of the present paper is to identify all the
possibilities for texture 6 zero as well as texture 5 zero
Fritzsch-like and non-Fritzsch like Hermitian quark mass matrices
which, in principle, are possible. Further, as a first step we
carefully examine these large number of possibilities, 144 for
texture 6 zero and 432 for texture 5 zero, to find out how many of
these represent independent Cabibbo-Kobayashi-Maskawa (CKM)
matrices \cite{ckm,ckm1}. Furthermore, these possibilities have been
examined keeping in mind the broad structure of CKM matrix, e.g.,
diagonal elements are near unity with non diagonal matrix elements
being very small. Finally, the reduced number of possibilities are
examined with a view to checking their compatibility with the
latest quark mixing data and by giving full variation to input
parameters as well as phases in the mass matrices.

To begin with, we first consider a typical Fritzsch-like texture
specific hermitian quark mass matrices as
 \be
 M_{U}=\left( \ba{ccc}
0 & A _{U} & 0      \\ A_{U}^{*} & D_{U} &  B_{U}     \\
 0 &     B_{U}^{*}  &  C_{U} \ea \right), \qquad
M_{D}=\left( \ba{ccc} 0 & A _{D} & 0      \\ A_{D}^{*} &
D_{D} &  B_{D}     \\
 0 &     B_{D}^{*}  &  C_{D} \ea \right),
\label{2zero}\ee
 where $M_{U}$ and $M_{D}$ correspond to up and
down mass matrices respectively . The Fritzsch-like texture 6 zero
matrices can be obtained from the above
 mentioned matrices by
taking both $D_{U}$ and $D_{D}$ to be zero, which reduces the
matrices $M_{U}$ and $M_{D}$ each to texture 3-zero. Fritzsch-like
texture 5 zero matrices can be obtained by taking either $D_{U}$ =
0 and $D_{D} \neq 0$ or $D_{U} \neq 0$ and $D_{D}$=0, thereby
giving rise to two possible cases of texture 5 zero matrices,
referred to as texture 5 zero $D_{U}$ = 0 pertaining to $M_{U}$
texture 3 zero type and $M_{D}$ texture 2 zero type and texture 5
zero $D_{D}$ = 0 case pertaining to $M_{U}$ texture
 2 zero type and $M_{D}$
texture 3 zero type. The non Fritzsch-like matrices differ from
the above mentioned Fritzsch-like matrices in regard to the
position of `zeros' in the structure of the mass matrices. One can
get non Fritzsch-like  mass matrices by shifting the position of
$C_i$ ($i=U, D$) on the diagonal as well as by shifting the
position of zeros among the non diagonal elements. For example, a
non Fritzsch-like texture 3 zero matrix is obtained if 1 $\times$
1 element is non zero, the other diagonal elements are zero
leaving the non diagonal elements unchanged. Similarly, by
considering 1 $\times$ 2 and 2 $\times$ 1 element to be zero and 1
$\times$ 3 and 3 $\times$ 1 element to be non zero, without
disturbing other elements, we again get texture 3 zero non
Fritzsch-like matrix. In a similar manner, one can obtain texture
2 zero non Fritzsch-like mass matrices by shifting the position of
zeros.

To discuss all possible combinations of texture 6 zero hermitian
mass matrices, we first identify the possible
patterns of texture 3 zero mass matrices. Before counting all
possibilities one has to keep in mind that these matrices have to
be diagonalized to yield quark masses as eigenvalues ($ m_1, -m_2,
m_3$), therefore we can conveniently impose the following
conditions
\be
{\rm Trace}~ M_{U,D}  \neq 0 \qquad {\rm and} \qquad {\rm Det}~
M_{U,D} \neq 0, \label{trace}\ee
 taking the second eigenvalue as $-m_2$ facilitates the
 diagonalization procedure without affecting the consequences \cite{fritzsch}.
 In view of non zero masses of
quarks, each of the possible texture pattern has to satisfy the
above condition. One can easily check that in case of texture 3
zero mass matrices we arrive at 20 different possible texture
patterns,
 out of which 8 are easily ruled out by imposing conditions mentioned
 in equation(\ref{trace}).
  To constrain these possible patterns of texture specific mass
matrices further, we consider constraints imposed by
diagonalization procedure of mass matrices in up and down sector
to obtain CKM matrix. Details of diagonalization
 can be looked up in our earlier work \cite{t6,t61,Monika,Monika1}.
An essential step in this process is to consider the invariants trace $M$,
trace $M^{2}$ and determinant $M$ which yield the relations  involving
elements of mass matrices and masses $m_1$, $-m_2$ and $m_3$
\cite{fritzsch,Monika,Monika1}. It is interesting to note that after
imposing conditions given in equation(\ref{trace}), 12 possible
textures break into two classes as shown in table(\ref{t1})
depending upon the equations these matrices satisfy. For example,
six matrices of class I, mentioned in table(\ref{t1}), satisfy the
following equations
\be
C = m_1- m_2+ m_3, \quad A^2 + B^2 = m_1m_2 + m_2m_3 -m_1m_3,
\quad A^2 C = m_1m_2m_3. \label{classI}\ee
 Similarly, in case of class II
all six matrices satisfy the following equations
\be
C + D= m_1 -m_2 +m_3,\quad A^2 -C D = m_1m_2 + m_2m_3 - m_1m_3,
\quad A^2 C = m_1m_2m_3. \label{classII}\ee
 The subscripts U and D have not been
used as these are valid for both kind of mass matrices.

After having classified these 12 possibilities into 2 distinct
classes one would like to find out the number of possible
combinations of $M_{U}$ and $M_{D}$ which yield distinct CKM
matrices. Matrices $M_{U}$ and $M_{D}$ each can correspond to any
of the 12 possibilities, therefore yielding 144 possible
combinations which in principle can yield 144 quark mixing
matrices. These 144 combinations can be put into 4 different
categories, e.g., if $M_{U}$ is any of the 6 matrices from class
I, then $M_{D}$ can be either from class I or class II yielding 2
categories of 36 matrices each. Similarly, we obtain 2 more
categories of 36 matrices each when $M_{U}$ is from class II and
$M_{D}$ is either from class I or class II. The 36 combinations in
each category further can be shown to be reduced to groups of six
combinations of mass matrices, each yielding same CKM matrix. For
example, six of the 36 combinations belonging to first category,
when both up and down sectors mass matrices are of the form
$M_{U_{i}}$ and $M_{D_{i}}$ (i = a,b,c,d,e,f), yield the same CKM
matrices. Similarly, the remaining 30 matrices in category one
yield five groups of six matrices each corresponding to five
independent mixing matrices. A similar simplification can be
achieved in other three categories.

The analysis get further simplified by using the fact that CKM
matrices depict clear hierarchy, i.e, the diagonal elements are
almost equal to unity and off diagonal elements are much smaller
than unity. Keeping these things in mind an analytical analysis of
categories as mentioned above yields only 4 groups of $M_{U_{i}}$
and $M_{D_{i}}$ corresponding to 6 combinations each of which
yield 4 CKM matrices showing hierarchical nature. To illustrate
this point we consider first matrix $M_{U}$ to be of type (a) from
class I and similarly $M_{D}$ to be of type (b) from the same
class. The corresponding CKM matrix is expressed as \beqn
 V_{{\rm CKM}} &=& \left( \ba {lll}
V_{ud} & V_{us} & V_{ub} \\ V_{cd} & V_{cs} & V_{cb} \\ V_{td} &
V_{ts} & V_{tb} \\ \ea \right)  \label{km0}  \eeqn

\be
=\left[ \ba {ccc}  -e^{i(\alpha_U - \alpha_D)} &
(\sqrt{\frac{m_d}{m_s}}) e^{i(\alpha_U - \alpha_D)} &
\sqrt{\frac{m_u}{m_c}}e^{i\beta_D}  \\
\sqrt{\frac{m_u}{m_c}}e^{i(\alpha_U - \alpha_D)}
+\sqrt{\frac{m_d}{m_b}}e^{i\beta_D} &
  -\sqrt{\frac{m_s}{m_b}} e^{i\beta_D}-\sqrt{\frac{m_c}{m_t}} e^{i\beta_U} &
e^{-i\beta_D}\\ \sqrt{\frac{m_d}{m_s}}e^{-i\beta_U}  & -1 &
\sqrt{\frac{m_s}{m_b}}e^{-i\beta_U} + \sqrt{\frac{m_c}{m_t}}
e^{-i\beta_D}   \ea \right], \label{np}\ee
 where $\alpha_i$ and
$\beta_i$, $i=U,D$ are related to the phases of the elements $A_i$
and $B_i$ of the mass matrices given in equation(\ref{2zero}). It
may be noted that in the above matrix, we have written the leading
terms only and used the hierarchy of quark mass matrices. From the
above structure of CKM matrix one can easily find out that off
diagonal elements, e.g., $|V_{cb}|$ and $|V_{ts}|$ are of the
order of unity whereas diagonal elements $|V_{cs}|$ and $|V_{tb}|$
are smaller than unity which is in complete contrast to the
structure of CKM matrix. In a similar manner, one can conclude
that remaining indistinguishable combinations also lead to such
non-physical mixing matrices. In case we apply the above criteria,
interestingly we are left with only four groups of mass matrices
as mentioned in table(\ref{t3}). Thus the problem of exploring the
compatibility of 144 phenomenological allowed texture 6 zero
combinations with the recent low energy data is reduced only to an
examination of 4 groups each having 6 combinations of mass
matrices corresponding to the same CKM matrix.

Considering now the case of texture 5 zero mass matrices which consist
either of $M_{U}$ being 2 zero and $M_{D}$ being 3 zero or vice versa.
 Texture 3 zero
possibilities have already been enumerated, therefore we consider only
the possible patterns of texture 2 zero mass matrices. After taking into
 consideration the condition mentioned in equation(\ref{trace}) one can
 check that there are 18
possible texture 2 zero patterns. These textures further break
into three classes detailed in table 2, depending upon the
diagonalization equations satisfied by these matrices, however, it
can be shown that the class IV and V essentially reduce to texture
3 zero patterns. For example, it can be easily shown that while
constructing the CKM matrix, the element F in type(a) matrix of
class V is much smaller
 than the other elements of considered mass matrix, therefore
it can be considered as a very small perturbation on corresponding
texture 3 zero pattern. Similar conclusion can be drawn for other
matrices of the same class as well as it can be shown that
matrices of class IV also reduce to texture 3 zero patterns.
Therefore, we conclude that the matrices in class IV and V
effectively reduce to the corresponding 3 zero patterns.  Hence,
there is no need to
 analyze these classes of texture 2 zero in the context of
texture 5 zero combinations. We are therefore left with only
one, class III of texture 2 zero, that needs to be explored
for texture 5 zero combinations.
All matrices of Class III satisfy the following equations
\be
 C + D = m_1- m_2+ m_3,\quad A^2 + B^2 - C D = m_1m_2 + m_2m_3 -m_1m_3,\quad A^2 C = m_1m_2m_3.
\label{classIII}\ee

Considering class III of texture 2 zero mass matrices along
with different patterns of class I and class II of texture
3 zero mass matrices we find a total of 144 possibilities of
 texture 5 zero mass matrices, in sharp contrast to the case
if we had considered class IV and class V also yielding 432
possibilities.
 Analysis similar to texture 6 zero
have been carried out for texture 5 zero mass matrices. Keeping in
mind the hierarchy of CKM matrix, we observe that out of 144
cases, we are again left with only 4 such groups of texture 5 zero
mass matrices detailed in table(\ref{t4}), leading to mixing
matrix having hierarchical structure as that of CKM matrix.

After having left with 4 groups each for texture 6 zero and 5 zero
mass matrices, we would like to check their compatibility with the
quark mixing data. To this end, we first consider the various
inputs used in the analysis. The quark masses and mass ratios,
considered at $M_z$ scale(GeV) \cite{mass}, are
\be
m_{u}=0.002- 0.003, m_{c}=0.6- 0.7, m_{t}=169.5- 175.5,
\label{uct}\ee
\be
 m_{d}=0.0037- 0.0052, m_{s}= 0.072- 0.097, m_{b}= 2.8- 3.0,
\label{dsb}\ee
\be
 m_{u}/m_{d} = 0.51- 0.60, m_{s}/m_{d} = 18.1-19.7 .
\label{ratio}\ee
 For the purpose of our calculations, we are
giving full variation to phases $\phi_{1}$ and $\phi_{2}$ and in
the case of texture 5 zero matrices the variation of the parameter
D is carried from 0 to $D<C$, as is usually done in such analyses.
Some of the CKM parameters so reproduced are to be compared with
the data taken from PDG (2008) \cite{pdg}, e.g.,
\be
|V_{us}| = 0.2236-0.2274, |V_{cb}|= 0.0401- 0.0423,
\label{vckm}\ee
\be
sin2\beta=0.656 -0.706,~~ J=(2.85-3.24)\times 10^{-5},~~
\delta=45^{\circ} - 107^{\circ}, \label{sin}\ee
 where $\beta$
represents the angle of the unitarity triangle and $J$ represents
the Jarlskog's rephasing invariant parameter with $\delta$ being
the CP violating phase of the CKM matrix.

Coming to the analysis of texture 6 zero mass matrices, in
table(\ref{t3}) we have presented the results of our analysis for
the four groups of texture 6 zero mass matrices each corresponding
to six combinations of $M_{U}$ and $M_{D}$. To check the viability
of these combinations, as a first step of our analysis, we first
reproduce $|V_{us}|$ element and then other elements of CKM
matrix. From the table one can immediately find that all possible
combinations of texture 6 zero are ruled out as these are not able
to reproduce the CKM element $|V_{cb}|$. Therefore, in the table
we have not given other CKM elements, however we have presented
the values of $sin2\beta$, $\delta$ and $J$. Thus, none of the
texture 6 zero combination, Fritzsch-like as well as non
Fritzsch-like, is found to be compatible with the recent quark
mixing data, ruling out the existence of these mass matrices.

In the case of texture 5 zero mass matrices, again we first
consider $|V_{us}|$ as a constraint and then try to reproduce
$|V_{cb}|$ and other elements of CKM matrix. In table(\ref{t4}),
we have presented the output of our analysis pertaining
 to texture 5 zero possibilities. A general look at the table immediately
 reveals that as compared to texture 6 zero case $|V_{cb}|$
predictions are quite different. However, interestingly only in
one case we are able to reproduce $|V_{cb}|$ as well as other CKM
elements in the desired range given by PDG 2008. Interestingly,
this possibility corresponds to usual Fritzsch-like texture 5 zero
mass matrix where $M_{U}$ is of texture 2 zero and $M_{D}$ is of
texture 3 zero. One may wonder that it looks somewhat strange
 that texture 5 zero matrix with $M_{U}$ being texture 2 zero and
$M_{D}$ being texture 3 zero is viable in contrast to the case
where $M_{U}$ being texture 3 zero and $M_{D}$ being texture 2
zero is not viable. This, however, can be easily understood if
one examines carefully the failure of texture 6 zero mass matrices
 to fit the data. In texture 6 zero case the hierarchy of quark masses
 gets translated to the hierarchy of elements of mass matrices therefore,
 it becomes difficult to fit the data. It becomes clear that
in order to fit the data we have to consider mass matrices where
 elements do not follow strong hierarchy of quark masses. In this
context one notes that hierarchy of quark masses is much stronger
 in up sector than in down sector. In case of texture 5 zero mass
 matrices when we consider up sector being 2 zero, then the
hierarchy of the corresponding mass matrix gets somewhat moderated
because of extra parameter making the present case viable. In
other case when $M_{D}$ is 2 zero and  $M_{U}$ continues to be 3
zero the hierarchy of elements of mass matrices comes out be very
strong, therefore it becomes difficult to fit the data.

\begin{figure}
\psfig{file=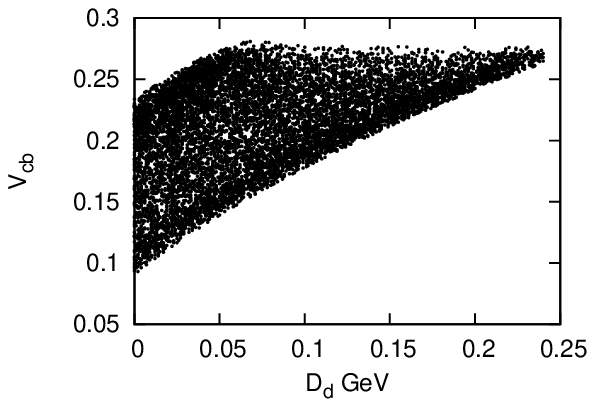, width=3.5in} \caption{Plot showing the
dependence of $|V_{cb}|$ on $D_D$}
  \label{vcbdd}
  \end{figure}

\begin{figure}
\psfig{file=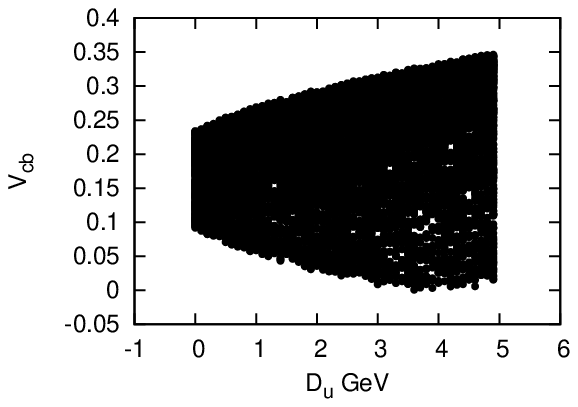, width=3.5in} \caption{Plot showing the
dependence of $|V_{cb}|$ on $D_U$}
  \label{vcbdu}
  \end{figure}

Interestingly the viable possibility mentioned above is able to
have a good overlap with experimental values of $|V_{cb}|$,
$|V_{ub}|$, $sin2\beta$ and $\delta$ \cite{pdg}. This can be
understood in a better way by investigating the dependence of CKM
matrix element $|V_{cb}|$ on $D_{U}$ and $D_{D}$ for the texture 5
zero cases. To this end, in figure \ref{vcbdd} we have plotted
$|V_{cb}|$ and $D_{D}$ corresponding to the texture 5 zero case
where $D_{U}=0$. From the figure it becomes clear that in no
situation we are able to reproduce $|V_{cb}|$. This is to be
contrasted with the texture 5 zero case where $D_{U} \neq 0$.
Again, in figure \ref{vcbdu} we have plotted the dependence of
$|V_{cb}|$ on $D_{U}$. As is evident from the figure, we are able
to reproduce $|V_{cb}|$ for $D_{U} \gtrsim 2$GeV, justifying the
above mentioned discussion.

\begin{figure}
\psfig{file=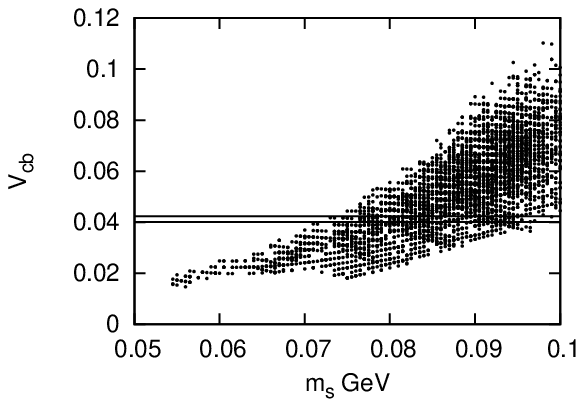, width=3.5in} \caption{Plot showing the
allowed range of $|V_{cb}|$ versus $m_s$}
  \label{vcbms}
  \end{figure}

\begin{figure}
\psfig{file=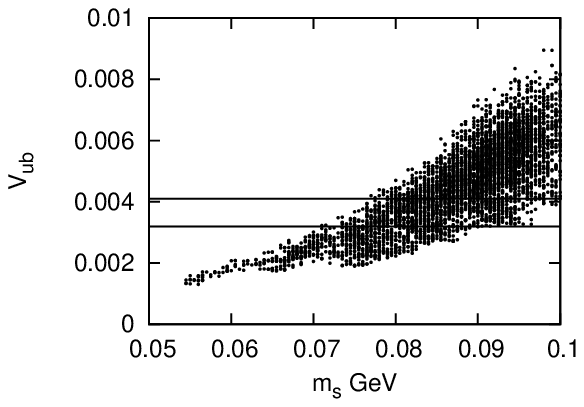, width=3.5in} \caption{Plot showing the
allowed range of $|V_{ub}|$ versus $m_s$}
  \label{vubms}
  \end{figure}

It should also be noted that agreement pertaining to texture 5
zero case mentioned above is valid only when Leutwyler \cite{mass}
quark masses are used. This agreement gets ruled out in case we
use the latest quark masses proposed by Xing et.al \cite{Xing
masses}. This can also be understood from a study of the
dependence of the elements $|V_{cb}|$ and $|V_{ub}|$ on the mass
$m_s$, shown in figures \ref{vcbms} and \ref{vubms}. While
plotting these figures, all other parameters have been given full
variation within the allowed ranges. From the figures, it
immediately becomes clear that when $m_s$ is lower than 0.075,
then one is not able to obtain both $|V_{cb}|$ and $|V_{ub}|$ in
the allowed range. Thus, one may conclude that viability of
texture 5 zero Fritzsch-like case is very much dependent on light
quark masses used.

To summarize, an extensive analysis of Fritzsch-like as well as
non-Fritzsch like texture 6 zero and 5 zero mass matrices have
been carried out. In all there are 144 Fritzsch-like as well as
non-Fritzsch like possibilities
 for texture 6 zero mass matrices whereas in case of texture 5 zero mass
matrices we have analyzed 432 possibilities. In the case of
texture 6 zero mass matrices these possibilities can be reduced to
24 possibilities which further break into groups of 6
possibilities each corresponding to the same CKM matrix.
Similarly, in the case of texture 5 zero mass matrices, the 432
possibilities interestingly again reduce to 4 groups each
corresponding to 6 possibilities with the same CKM matrix.
Surprisingly, all the texture 6 zero possibilities are completely
ruled out whereas in case of texture 5 zero mass matrices there is
one unique Fritzsch-like combination which shows limited viability
depending upon the light quark masses used as input, the other
possibilities are again ruled out.

 {\bf ACKNOWLEDGMENTS}\\ The authors would like to
thank Gulsheen Ahuja for going through the manuscript. N.M. would
like to thank Chairman, Department of Physics for providing
facilities to work in the department. R.V. would like to thank the
Director, RIEIT for providing facilities to work.

\begin{table}
\bt{|c|c|c|} \hline
  & Class I  & Class II \\ \hline
a & $\left ( \ba{ccc} {\bf 0} & Ae^{i\alpha} & {\bf 0} \\
Ae^{-i\alpha}  & {\bf 0} & Be^{i\beta} \\ {\bf 0} & Be^{-i\beta}
& C \ea \right )$  & $\left ( \ba{ccc} {\bf 0} & Ae^{i\alpha} &
{\bf 0} \\ Ae^{-i\alpha}  & D & {\bf 0} \\ {\bf 0} & {\bf 0}  & C
\ea \right )$ \\ b &  $\left ( \ba{ccc} {\bf 0} &{\bf 0} &
Ae^{i\alpha} \\ {\bf 0}  & C & Be^{i\beta} \\Ae^{-i\alpha} &
B^{-i\beta}  & {\bf 0} \ea \right )$  &
 $\left ( \ba{ccc} {\bf 0} & {\bf 0} & Ae^{i\alpha}
 \\ {\bf 0}  & C & {\bf 0} \\ Ae^{-i\alpha}  & {\bf 0}  &
D \ea \right )$ \\ c &  $\left ( \ba{ccc} {\bf 0} & Ae^{i\alpha} &
Be^{i\beta} \\ Ae^{-i\alpha}  & {\bf 0} & {\bf 0} \\ Be^{-i\beta}
& {\bf 0}  & C \ea \right )$  &
 $\left ( \ba{ccc} D & Ae^{i\alpha} &
{\bf 0} \\ Ae^{-i\alpha}  & {\bf 0} & {\bf 0} \\ {\bf 0} & {\bf 0}
& C \ea \right )$ \\ d &  $\left ( \ba{ccc} C & Be^{i\beta} & {\bf
0}
\\ Be^{-i\beta}  & {\bf 0} & Ae^{i\alpha}\\ {\bf 0}  & Ae^{-i\alpha} &
{\bf 0} \ea \right )$  &
 $\left ( \ba{ccc} C & {\bf 0} & {\bf 0}
 \\ {\bf 0}  & D & Ae^{i\alpha} \\ {\bf 0} & Ae^{-i\alpha}  &
{\bf 0} \ea \right )$ \\ e &  $\left ( \ba{ccc} {\bf 0} &
Be^{i\beta}  & Ae^{i\alpha} \\ Be^{-i\beta}  & C & {\bf 0} \\
Ae^{-i\alpha}  & {\bf 0}  &
 {\bf 0} \ea \right )$  &
 $\left ( \ba{ccc} D & {\bf 0} & Ae^{i\alpha}
\\ {\bf 0} & C &  {\bf 0} \\ Ae^{-i\alpha}  & {\bf 0}  &
{\bf 0}  \ea \right )$ \\ f & $\left ( \ba{ccc} C & {\bf 0} &
Be^{i\beta}
 \\ {\bf 0}  & {\bf 0}  & Ae^{i\alpha} \\Be^{-i\beta} & Ae^{-i\alpha}  &
{\bf 0} \ea \right )$  &
 $\left ( \ba{ccc} C & {\bf 0} &{\bf 0}
 \\ {\bf 0}  & {\bf 0} & Ae^{i\alpha} \\ {\bf 0} & Ae^{-i\alpha}  &
D \ea \right )$ \\    \hline \et \caption{Twelve possibilities of
texture 3 zero mass matrices categorized into two classes I and II
satisfying the conditions given in equation(\ref{trace}), with
each class having six matrices labeled as a,b,c,d,e,f. }\label{t1}
\end{table}
\newpage
\begin{table}
\bt{|c|c|c|c|} \hline
 & Class III  & Class IV  & Class V   \\  \hline
a & $\left ( \ba{ccc} {\bf 0} & Ae^{i\alpha} & {\bf 0} \\
Ae^{-i\alpha}  & D & Be^{i\beta} \\ {\bf 0} & Be^{-i\beta}  & C
\ea \right )$  &
 $\left ( \ba{ccc} D & Ae^{i\alpha} &
{\bf 0} \\ Ae^{-i\alpha}  & {\bf 0} &  Be^{i\beta} \\ {\bf 0} &
Be^{-i\beta}  & C \ea \right )$     &
 $\left ( \ba{ccc} {\bf 0} & Ae^{i\alpha} &
 Fe^{i\gamma}\\ Ae^{-i\alpha}  &{\bf 0}  & Be^{i\beta} \\ Fe^{-i\gamma} & Be^{-i\beta}  &
C \ea \right )$ \\ b &  $\left ( \ba{ccc} {\bf 0} & {\bf 0}  &
Ae^{i\alpha} \\ {\bf 0}  & C & Be^{i\beta} \\  Ae^{-i\alpha} &
Be^{-i\beta}  & D \ea \right )$  &
 $\left ( \ba{ccc} D & {\bf 0} & Ae^{i\alpha}
 \\ {\bf 0} & C & Be^{i\beta} \\  Ae^{-i\alpha} & Be^{-i\beta}  &
{\bf 0} \ea \right )$     & $\left ( \ba{ccc} {\bf 0} &
Fe^{i\gamma} & Ae^{i\alpha}
\\  Fe^{-i\gamma}  & C  & Be^{i\beta} \\  Ae^{-i\alpha} & Be^{-i\beta}  &
 {\bf 0}\ea \right )$ \\
c &  $\left ( \ba{ccc} D & Ae^{i\alpha} &
  Be^{i\beta}\\ Ae^{-i\alpha}  & {\bf 0}  & {\bf 0} \\  Be^{-i\beta} & {\bf 0}
& C \ea \right )$  & $\left ( \ba{ccc} {\bf 0} & Ae^{i\alpha} &
 Be^{i\beta} \\ Ae^{-i\alpha}  & D & {\bf 0}  \\  Be^{-i\beta}  & {\bf 0} &
C \ea \right )$     & $\left ( \ba{ccc} {\bf 0} & Ae^{i\alpha} &
Be^{i\beta}
\\ Ae^{-i\alpha} & {\bf 0} & Fe^{i\gamma} \\ Be^{-i\beta}  & Fe^{-i\gamma} &
C \ea \right )$ \\ d &  $\left ( \ba{ccc} C & Be^{i\beta} & {\bf
0}
 \\ Be^{-i\beta} & D & Ae^{i\alpha}  \\ {\bf 0} & Ae^{-i\alpha} & {\bf 0}
 \ea \right )$  &
 $\left ( \ba{ccc} C & Be^{i\beta} &  {\bf 0}
 \\ Be^{-i\beta} & {\bf 0} & Ae^{i\alpha} \\ {\bf 0} & Ae^{-i\alpha}  &
D \ea \right )$     &
 $\left ( \ba{ccc} C & Be^{i\beta}  & Fe^{i\gamma}
\\ Be^{-i\beta}   & {\bf 0}  &  Ae^{i\alpha} \\ Fe^{-i\gamma} & Ae^{-i\alpha}  &
 {\bf 0}\ea \right )$ \\
e &  $\left ( \ba{ccc} D & Be^{i\beta} & Ae^{i\alpha}
  \\ Be^{-i\beta}  & C & {\bf 0} \\ Ae^{-i\alpha} & {\bf 0}
& {\bf 0} \ea \right )$  &
 $\left ( \ba{ccc} {\bf 0} & Be^{i\beta} & Ae^{i\alpha}
  \\  Be^{-i\beta}  & C & {\bf 0}  \\ Ae^{-i\alpha}   & {\bf 0} &
D \ea \right )$     & $\left ( \ba{ccc} {\bf 0}  & Be^{i\beta} &
Ae^{i\alpha}
\\ Be^{-i\beta}  & C & Fe^{i\gamma} \\ Ae^{-i\alpha} & Fe^{-i\gamma}&
{\bf 0}  \ea \right )$ \\ f &  $\left ( \ba{ccc} C & {\bf 0} &
Be^{i\beta}
 \\ {\bf 0} & {\bf 0} & Ae^{i\alpha}  \\ Be^{-i\beta} & Ae^{-i\alpha} & D
 \ea \right )$  &
$\left ( \ba{ccc} C  &  {\bf 0} & Be^{i\beta}
 \\ {\bf 0} & D & Ae^{i\alpha} \\Be^{-i\beta}  & Ae^{-i\alpha}  &
0 \ea \right )$     &
 $\left ( \ba{ccc} C & Fe^{i\gamma} & Be^{i\beta}
\\  Fe^{-i\gamma} & {\bf 0}  &  Ae^{i\alpha} \\ Be^{i\beta} & Ae^{-i\alpha}  &
 {\bf 0}\ea \right )$ \\  \hline
\et \caption{Texture 2 zero possibilities categorized into three
classes III, IV and V, satisfying the conditions given in equation
(\ref{trace}), with each class having six matrices labeled as
a,b,c,d,e,f.}\label{t2}
\end{table}

\begin{table}
\bt{|cc|c|c|c|c|c|} \hline $M_{U}$ & $M_{D}$ & $|V_{cb}|$ &
$sin2\beta$ & $\delta^{\circ}$ & $J\times10^{-5}$  \\ \hline
$I_{i}$ & $I_{i}$ & 0.09-0.24 & 0.48-0.57 & 78-100 & 12-76  \\
$II_{i}$ & $II_{i}$  & 0 & not defined & not defined & 0 \\
$I_{i}$ & $II_{i}$  & 0.055-0.065 & 0.48-0.55 & 74-90 & 4.7-5.3 \\
$II_{i}$ & $I_{i}$  & 0.15-0.18 & 0.50-0.54 & 80-95 & 32-44 \\
\hline \et \caption{Predicted values of $|V_{cb}|$,  $sin2\beta$,
$\delta$ and J for 4 independent texture 6 zero combinations where
i = a,b,c,d,e,f .} \label{t3}\end{table}

\begin{table}
\bt{|cc|c|c|c|c|c|c|} \hline $M_{U}$ & $M_{D}$  & $|V_{cb}|$ &
$|V_{ub}|$ & $sin2\beta$ & $\delta^{\circ}$ & $J\times10^{-5}$  \\
\hline $I_{i}$ & $III_{i}$ & 0.09-0.28 & 0.005-0.02 & 0.44-0.60 &
60-100 & 10-110 \\ $II_{i}$ & $III_{i}$  & 0.14-0.26 & 0.008-0.02
& 0.46-0.59 & 65-93 & 0.27-0.97 \\ $III_{i}$ & $I_{i}$  &
0.0401-0.0423 & 0.0032-0.0041 & 0.656-0.701 & 55-100 & 2.4-3.8 \\
$III_{i}$ & $II_{i}$  & 0.06-0.29 & 0.003-0.02 & 0.51-0.54 & 45-88
& 4.8-110   \\ \hline \et \caption{Texture 5 zero combinations and
their corresponding predicted values pertaining to $|V_{cb}|$,
$|V_{ub}|$, $sin2\beta$, $\delta$ and $J$.} \label{t4}
\end{table}

\newpage

\end{document}